\documentclass[
    ,final            
  ]
  {aipproc}

\layoutstyle{6x9}

\usepackage{txfonts}
\usepackage{natbib}
\bibpunct{(}{)}{;}{a}{}{,} 

\def\mso{\,\mathrm{M}_\odot}

 \def\simle{\mathrel{\hbox{\rlap{\hbox{\lower4pt\hbox{$\sim$}}}\hbox{$<$}}}}
 \def\simgr{\mathrel{\hbox{\rlap{\hbox{\lower4pt\hbox{$\sim$}}}\hbox{$>$}}}}

 \def\grad{\nabla}
 \def\adgrad{\nabla_{\mathrm{\!ad}}}
 \def\mugrad{\nabla_{\!\mu}}
 \def\ath{\alpha_{\mathrm{th}}}

 \def\c2{^{12}{\rm C}}
 \def\c3{^{13}{\rm C}}
 \def\n14{^{14}{\rm N}}
 \def\c1213{^{12}{\rm C}/^{13}{\rm C}}
 \def\he3he4{^3{\rm He}/^4{\rm He}}

\begin{document}

\title{Thermohaline mixing in low-mass giants:\\ RGB and beyond}

\classification{97.}
\keywords      {Stars: evolution -- Stars: rotation -- Stars: mixing processes }

\author{M.Cantiello}{
  address={Astronomical Institute, Utrecht University,\\
              Princetonplein 5, 3584 CC, Utrecht, The Netherlands}
}

\author{H.Hoekstra}{
 }

\author{N.Langer}{
 }
\author{A.J.T.Poelarends}{
 }

\begin{abstract}
Thermohaline mixing has recently been proposed to occur in
low mass red giants, with large consequence for the chemical yields
of low mass stars. 
We investigate the role of thermohaline mixing during the evolution
of stars between 1$\mso$ and 3$\mso$, in comparison to other mixing processes
acting in these stars.  
We use a stellar evolution code which includes rotational mixing and
internal magnetic fields.    
We confirm that thermohaline mixing has the potential to destroy most of
the $^3$He which is produced earlier on the main sequence during the red giant
stage, in stars below $1.5\mso$. We find this process to continue during core helium
burning and beyond. We find rotational and magnetic mixing to be negligible compared to
the thermohaline mixing in the relevant layers, even if the interaction of thermohaline 
motions with the differential rotation may be essential to establish the time scale
of thermohaline mixing in red giants.
 
\end{abstract}

\maketitle


\section{Introduction}
Thermohaline mixing is usually not considered as an important mixing process
in single stars, since the ashes of thermonuclear fusion consists of heavier nuclei
than its fuel, and stars usually burn from the inside out. The condition for
thermohaline mixing, however, is that the mean molecular weight ($\mu$)
decreases inward. 
Recently \citet[CZ07]{cz07} identified thermohaline mixing 
as an important mixing process which significantly modifies the surface composition
of red giants after the first dredge-up. The work by CZ07 was triggered by the paper
of \citet[EDL06]{edl06}, who found a mean molecular weight ($\mu$) 
inversion --- i.e., $\left( d\log\mu \over d\log P \right) < 0$ --- 
below the red giant convective envelope in a 1D-stellar evolution
calculation.

EDL06 found a $\mu$-inversion in their $1\mso$ stellar evolution model, occurring
after the so-called luminosity bump on the red giant branch, which is produced
after the first dredge-up, when the hydrogen-burning shell source enters the
chemically homogeneous part of the envelope. The $\mu$-inversion is produced by the
reaction $^3$He($^3$He,2p)$^4$He (as predicted by \citet{ulr72}). It does not show up earlier,
since the magnitude of the $\mu$-inversion is small, and negligible if compared to a stabilizing
$\mu$-stratification.

The mixing process below the convective envelope in models of low mass stars turns out
to be essential for the prediction of the chemical yield of $^3$He (EDL06); this process is 
also essential to understand the surface abundances of red giants, in particular the $^{12}$C/$^{13}$C 
ratio, the $^7$Li and the carbon and nitrogen
abundances (CZ07). We investigate the evolution
of solar metallicity stars between $1\mso$ and $3\mso$ from the ZAMS up to the
thermally-pulsing AGB stage, based on models computed 
during the last years. We show for which initial mass range, and during which evolutionary
phase thermohaline mixing occurs, and with which consequences. Besides thermohaline mixing,
our models include
convection, rotation-induced mixing, and internal magnetic fields, and we compare 
the significance of these processes in relation to the thermohaline mixing.

\section{Method}
We compute evolutionary models of 1.0, 1.5, 2.0 and $3.0\mso$ with solar metallicity (Z=0.02).
We use a hydrodynamic stellar evolution code which includes the effect of rotation and magnetic fields \citep[e.g.][]{hlw00,yl05b}.
Mixing of chemical species is treated as a diffusive process.
The condition for the occurrance of thermohaline mixing is
\begin{equation}
\frac{\varphi}{\delta} \,\mugrad \le \grad - \adgrad \le 0 \label{condition}
\end{equation}
i.e. the instability operates in regions that are stable against convection (according to the 
Ledoux criterion) and where an inversion in the mean molecular weight is present.
Numerically, we treat thermohaline mixing through a diffusion scheme
\citep{bra97,wlb01}. The corresponding diffusion coefficient is based on the
work of Stern (1960), \citet{ulr72}, and \citet{krt80}; it
reads
\begin{equation}
  D_{th} = -\ath \frac{3K}{2c_P\rho}
  \frac{\frac{\phi}{\delta}\nabla_\mu}{\nabla_{ad} - \nabla} \label{coefficient} 
\end{equation}
where $K=4acT^{3}/(3\kappa\rho )$, $\phi =(\partial \ln \rho / \partial \ln
\mu )_{P,T}$, $\delta =-(\partial \ln \rho / \partial \ln T )_{P,\mu
  }$, $\nabla_\mu = d\ln\mu / d\ln P$, $\nabla_{ad}= (\partial \ln T/
\partial \ln P )_{ad}$, and $\nabla = d\ln T / d\ln P$.  The quantity
$\ath$ is a efficiency parameter for the thermohaline mixing.
The value of this parameter 
depends on the geometry of the fingers arising from the instability and is still a matter of debate 
\citep{ulr72,krt80,cz07}. We use a value of $\ath$ corresponding to the prescription of \citet{krt80}.

 \begin{figure}
 \begin{minipage}{7.5cm}
  \resizebox{\hsize}{!}{\includegraphics{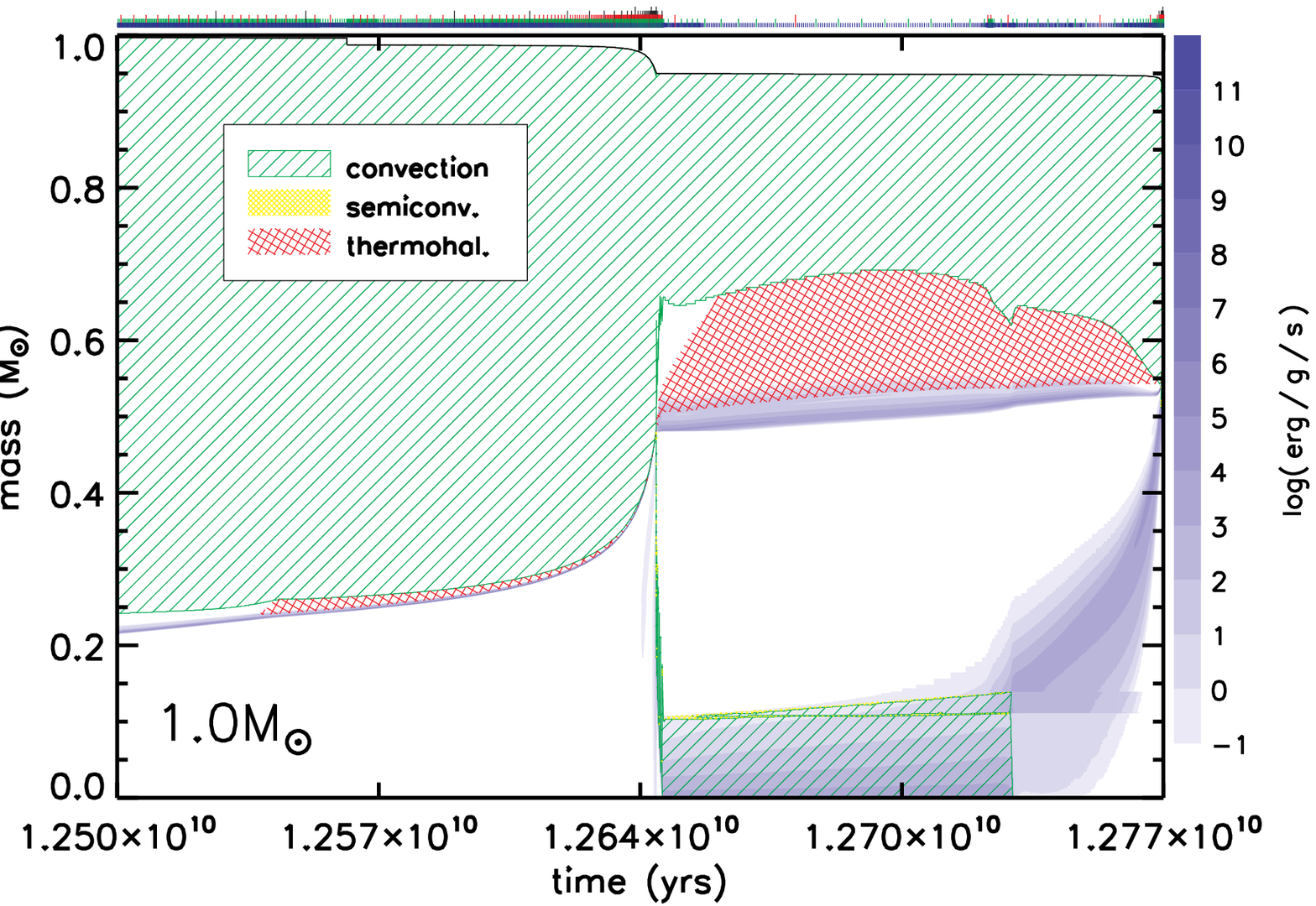}}
 \end{minipage}
 \hfill
 \begin{minipage}{7.5cm} 
   \resizebox{\hsize}{!}{\includegraphics{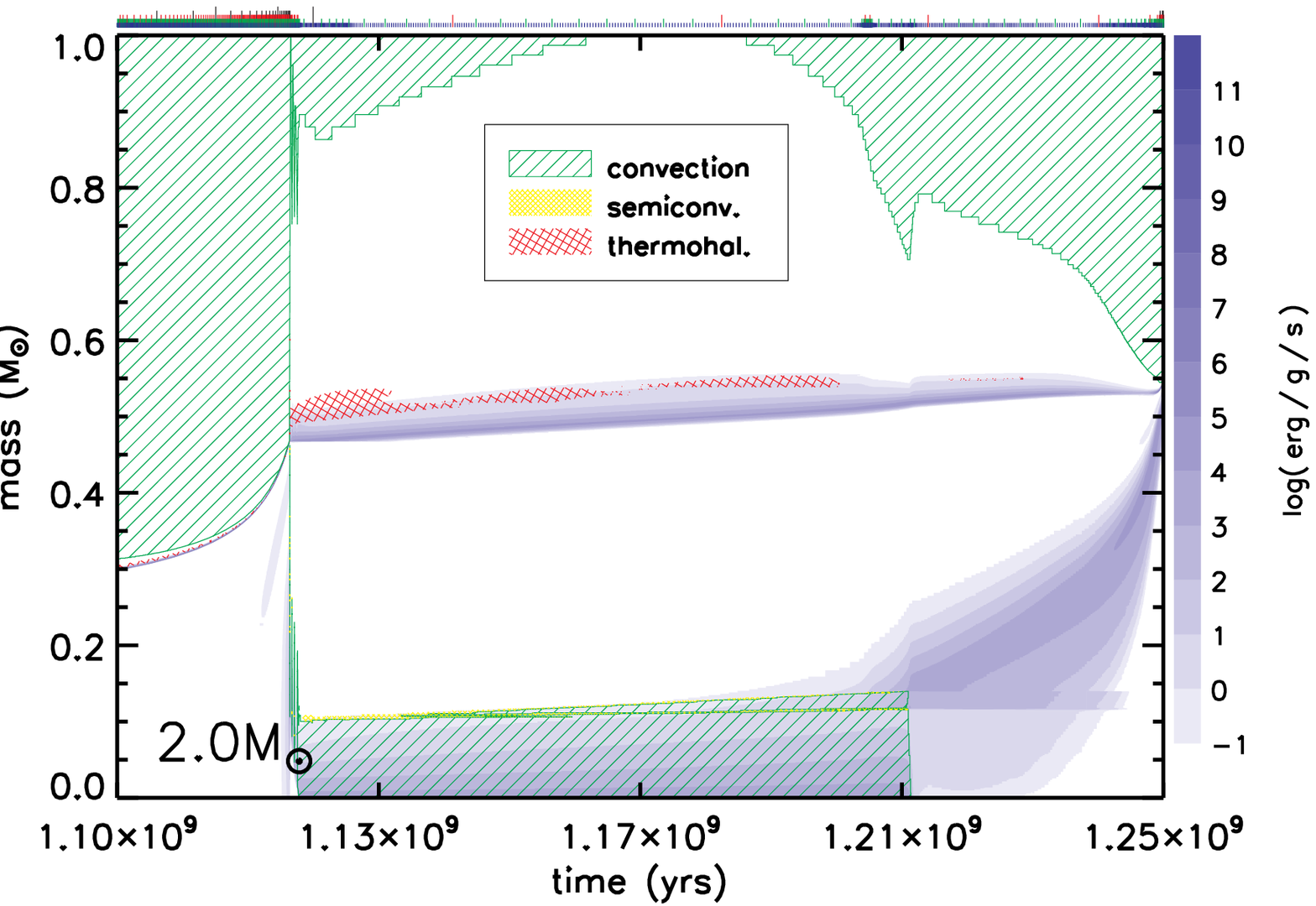}}
 \end{minipage}
\caption{{\bf Left panel:} evolution of the internal structure of a $1.0\mso$ star from the onset of thermohaline 
mixing to the AGB phase. Green hatched regions indicate convection, yellow
 filled regions represent semiconvection and red cross hatched regions indicate thermohaline mixing, 
as displayed in the legend. Blue shading shows regions of nuclear energy generation. {\bf Right panel:} same as left panel, but for a $2.0\mso$ star.}
\label{1e2}
 \end{figure}

\section{RGB and beyond}
The surface composition of low mass stars is substantially changed during the first dredge-up: lithium and 
carbon abundances as well as the carbon isotopic ratio decline,$^3$He and nitrogen abundances increase.
 After the first dredge-up the hydrogen-burning shell is advancing while the convective envelope retreats;
 the shell source then  enters the chemically homogeneous part of the envelope. EDL06 and CZ07  have shown 
 how in this situation an inversion in the molecular weight is created by
 the  reaction $^3$He($^3$He,2p)$^4$He in the outer wing of the hydrogen-burning shell in  models of 1.0 
and 0.9 $\mso$. This inversion is responsible for thermohaline mixing to develop. 

\begin{figure}
 \begin{minipage}{7.5cm}
  \resizebox{\hsize}{!}{\includegraphics{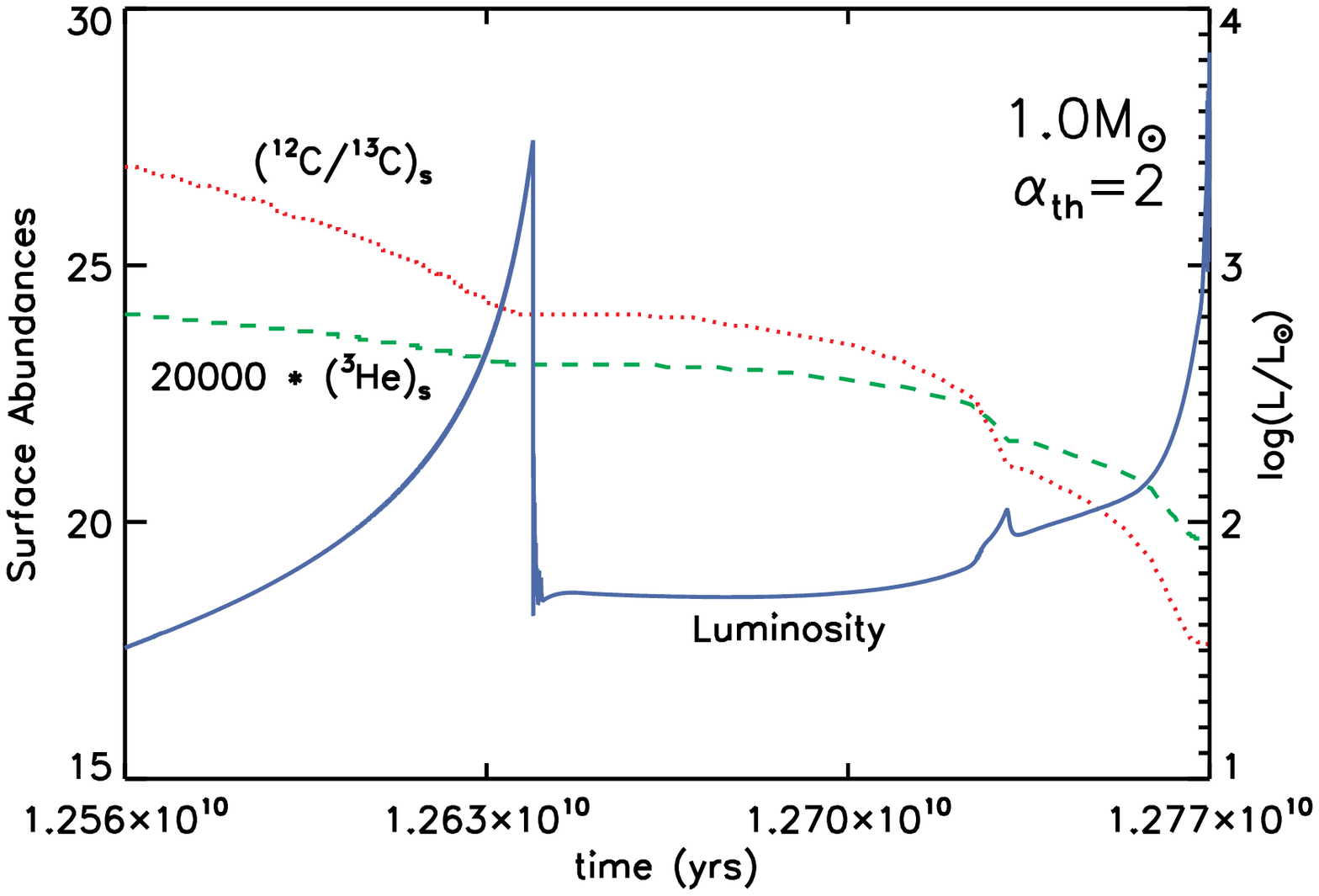}}
 \end{minipage}
 \hfill
 \begin{minipage}{7.5cm} 
   \resizebox{\hsize}{!}{\includegraphics{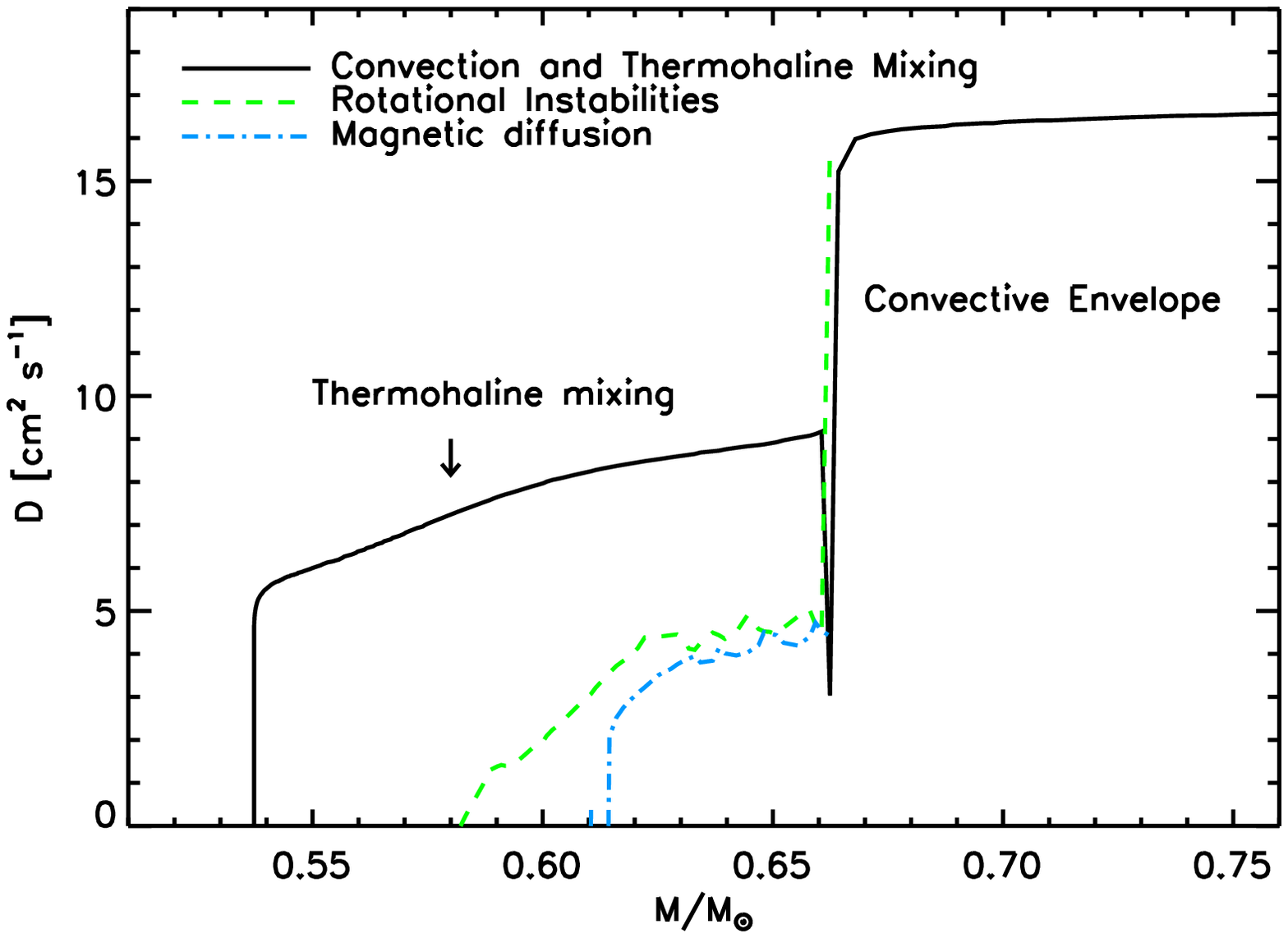}}
 \end{minipage}
\caption{{\bf Left panel:} evolution of the surface abundance of the $^{12}$C/$^{13}$C ratio 
(dotted red line) and $^3$He (dashed green line), and of the luminosity (solid blue line)
 from the onset of thermohaline mixing up to the AGB phase for a $1.0\mso$ star. {\bf Right panel:} diffusion coefficients 
in the region between the H burning shell and the convective envelope 
for the $1.0\mso$ model during the RGB phase (t $=1.267\times10^{10}$ years). 
The black, continuous line shows convective and thermohaline mixing diffusion coefficients, the green, 
dashed line is the sum of the diffusion coefficients due to rotational 
instabilities while the blue, dot-dashed line shows the magnitude of magnetic diffusion coefficient.}
\label{surfacediff}
 \end{figure}

We compute stellar models of 1.0, 1.5, 2.0 and 3.0 $\mso$ with solar metallicity including the effects of
rotation and magnetic fields. We confirm the presence of an inversion in the mean molecular weight, in
the outer wing of the H-burning shell, after the luminosity bump on the red giant branch.
According to inequality (\ref{condition}) this inversion gives rise to  thermohaline mixing  in the radiative 
buffer layer, the radiative region between the H-burning shell and the convective envelope.\\
In our 1$\mso$ model, thermohaline mixing developes at the luminosity bump and transports chemical species in  
the radiative layer between the H-burning shell and the convective envelope. This results 
in a change of the stellar surface abundances. The left panel of Fig. \ref{surfacediff} shows the evolution of $^3$He surface abundance 
and of the ratio $\c1213$ at surface as a function of time, confirming the result of EDL06 and CZ07, namely that thermohaline mixing
is efficiently depleting $^3$He and lowering the ratio $\c1213$ on the giant branch.

While CZ07 and EDL07 investigate thermohaline mixing only during the RGB phase, we followed the evolution of 
our models until the TP-AGB phase.
Infact a $\mu$-inversion is always created if a H-burning shell is active in a chemically homogeneous layer; this  happens not only during the RGB phase, 
but also during the HB and AGB phases. The size of the $\mu$-inversion  is depending on the local amount of $^3$He, that comes from the incomplete PP chain,
 as well as from
the chemically homogeneous layer.  

After core He-flash, helium is burned in the core, while a H-burning shell is still active below the convective envelope.
We found that during this phase thermohaline mixing is present and can spread through the whole radiative buffer layer in our 1$\mso$ model (left panel in Fig. \ref{1e2}). 
In this model the surface abundances change also during this phase because the H-burning shell and the envelope are connected. This is shown in  Fig. 
\ref{surfacediff}, left panel, where surface abundances change also after the luminosity peak corresponding to the He-flash.
 We stress that using the prescription of \citet{krt80} for thermohaline mixing allows our model to reach this phase without completely burning the $^3$He; 
models of CZ07 almost completely 
deplete $^3$He in the envelope already during 
the RGB phase because of their higher diffusion coefficient.
In this case thermohaline mixing would be much less efficient, during the subsequent evolutionary phases, due to the lower abundance of $^3$He.

\begin{figure}
\includegraphics[height=.3\textheight]{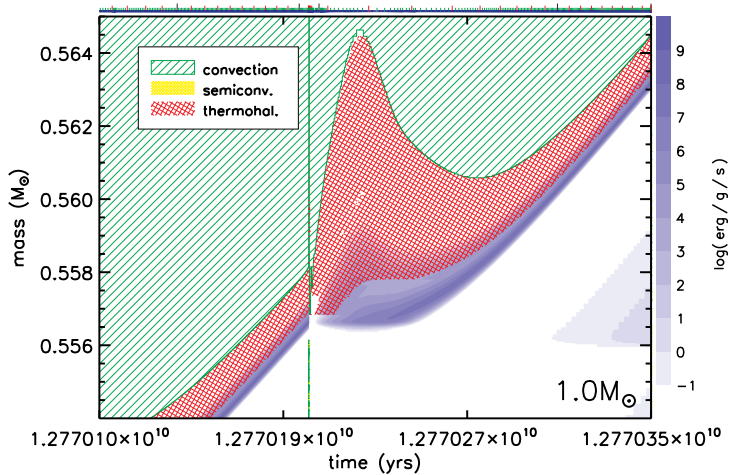}
 \caption{Evolution of 
the region between the H burning shell source and the convective 
envelope during a thermal pulse in a $1.0\mso$ star. Green hatched regions indicate convection and red 
crossed regions indicate thermohaline mixing. Blue shading shows regions of nuclear energy generation. }
\label{pulse} 
\end{figure}

The subsequent evolutionary phase of a low mass star is referred as Asymptotic Giant Branch (AGB), and is characterized by the presence of two burning shells and a degenerate core.
The star burns H in a shell and the ashes of this process feed a underlying He shell. 
During the most luminous part of the AGB the He shell periodically experiences thermal pulses (TPs); in stars more massive
than 2$\mso$ these thermal pulses are associated with a deep penetration of the convective envelope, the so-called third dredge-up (3DUP).
We find thermohaline mixing to be present also in the TP-AGB phase. Depending on the mass of the model the diffusion process is able to connect the
 H-burning shell with the convective envelope during the whole interpulse phase.
In a $1\mso$ model thermohaline mixing connects the H-burning shell 
to the
convective envelope (Fig. \ref{pulse}), confirming that this mixing process is more efficient at lower masses.

\section{Rotation and magnetic fields}
In our models we found that in the relevant layers thermohaline mixing
has generally higher diffusion coefficients than rotational instabilities and magnetic diffusion. 
The right panel of Fig. \ref{surfacediff} clearly shows that rotational and magnetic mixing are negligible compared to
the thermohaline mixing in our $1.0\mso$ model. The only rotational instability acting on a shorter
timescale is the dynamical shear instability, visible in the right panel of Fig. \ref{surfacediff} as a spike present at  the lower 
boundary of the convective envelope.
This instability works on the dynamical timescale in regions of a star where a high degree of differential rotation is present; it 
sets in if the energy that can be gained from the shear flow becomes comparable to the work which has to be done against the
potential for an adiabatic turn-over of a mass element (``eddy'') \citep{heg98}. However, if present, this instability acts only in a 
very small region (in mass coordinate) at the bottom of the convective envelope. As a result thermohaline mixing is still setting the
timescale for the diffusion of chemical species from the convective envelope to the Hydrogen-burning shell.

We will discuss qualitatively the interaction of thermohaline 
motions with magneto-rotational instabilities in a forthcoming paper \citep{chl+07}.

\section{Discussion}
We comfirm the results of EDL06 and CL07: thermohaline mixing in low mass giants is capable
of destroying large quantities of ${^3}$He, as well as decreasing the ratio $\c1213$.
Thermohaline mixing indeed starts when the hydrogen burning
shell source moves into the chemically homogeneous layers established by the first dredge-up. 
Our models show further that thermohaline mixing remains important during core helium burning,
and can still be relevant during the AGB phase --- including the termally-pulsing AGB stage.   
This results in important changes in the surface abundances of low mass stars. The quantitative discussion is complicated
by the fact that thermohaline mixing is strongly dependent on the mass of the star and on the efficiency of thermohaline mixing, which is still a matter of debate.

Moreover, our calculations show that in the relevant layers thermohaline mixing has generally a higher diffusion coefficient than rotational instabilities
and magnetic diffusion. 

We will discuss qualitatively the interaction of thermohaline mixing with magneto-rotational instabilities in a forthcoming paper, where we will also explore
the effect of using different prescriptions for thermohaline mixing diffusion coefficient \citep{chl+07}.


  


\bibliographystyle{mn2e}      

\bibliography{haline}

\IfFileExists{\jobname.bbl}{}
 {\typeout{}
  \typeout{******************************************}
  \typeout{** Please run "bibtex \jobname" to optain}
  \typeout{** the bibliography and then re-run LaTeX}
  \typeout{** twice to fix the references!}
  \typeout{******************************************}
  \typeout{}
 }

\end{document}